\documentclass{article}
\usepackage{ijcai19}

\usepackage{times}
\usepackage{soul}
\usepackage{url}
\usepackage[hidelinks]{hyperref}
\usepackage[utf8]{inputenc}
\usepackage[small]{caption}
\usepackage{graphicx}
\usepackage{amsmath}
\usepackage{booktabs}
\usepackage{algorithm}
\usepackage{algorithmic}
\title{MUTLA: A Large-Scale Dataset for Multimodal Teaching and Learning Analytics}

 \author{
 Fangli Xu$^1$\footnote{Corresponding Author}\and
 Lingfei Wu$^2$\and
 KP Thai$^1$\and
 Carol Hsu $^3$ \and
 Wei Wang$^1$\and
 Richard Tong$^1$
 \affiliations
 $^1$Squirrel AI Learning by Yixue Education Inc., New Jersey \\
 $^2$IBM T.J. Waston Research Center, Yorktown Heights, New York\\
 $^3$ Songshu AI, Shanghai, China
 \emails
 \{kp.thai, wei, richard\}@yixue.us, 
 fanglixu55@gmail.com, lwu@email.wm.edu, xuzhaohui@songshuai.com, 
 }

\begin{document}

\maketitle
\begin{abstract}
  Automatic analysis of teacher and student interactions could be very important to improve the quality of teaching and student engagement. However, despite some recent progress in utilizing multimodal data for teaching and learning analytics, a thorough analysis of a rich multimodal dataset coming for a complex real learning environment has yet to be done. To bridge this gap, we present a large-scale MUlti-modal Teaching and Learning Analytics (MUTLA) dataset. This dataset includes time-synchronized multimodal data records of students (learning logs, videos, EEG brainwaves) as they work in various subjects from Squirrel AI Learning System (SAIL) to solve problems of varying difficulty levels. The dataset resources include user records from the learner records store of SAIL, brainwave data collected by EEG headset devices, and video data captured by web cameras while students worked in the SAIL products. Our hope is that by analyzing real-world student learning activities, facial expressions, and brainwave patterns, researchers can better predict engagement, which can then be used to improve adaptive learning selection and student learning outcomes. An additional goal is to provide a dataset gathered from the real-world educational activities versus those from controlled lab environments to benefit educational learning community. 
\end{abstract}

\section{Introduction}

Recent in advancements in Artificial Intelligence Deep Learning \cite{lecun2015deep} have generated fast-growing interests in applying advanced machine learning/deep learning techniques to education. In particular, there has been substantial interest in multimodal teaching and learning analytics (MUTLA), a research methodology that aims to bring together Educational Data Mining and Learning Analytics to multimodal learning environments by directly working on  data from multimodalities \cite{worsley2018multimodal,worsley2016situating}. Over last ten years, researchers have exploited to apply machine learning/ data mining techniques on multimodal data for various tasks, including communicative interaction \cite{macwhinney2004talkbank}, online education \cite{thomas2018multimodal}, student’s uncertainty modeling \cite{jraidi2013student}, and emotional responses recognition in children \cite{nojavanasghari2016emoreact}.

Despite some recent progress of collecting multimodal data and utilizing it in learning science \cite{macwhinney2004talkbank,antoniadou2017collecting,oviatt2013multimodal,nojavanasghari2016emoreact}, teaching and learning analytics is largely limited by the quality and quantity of multimodal data that is publicly accessible. So here in this paper, we bridge this gap between enthusiastic AI researchers and challenging teaching and learning analytics problems, by presenting a large-scale and real-world MUTLA dataset that is publicly accessible.

Our dataset includes time-synchronized multimodal data recordings (learning logs, videos, EEG brainwaves) on students as they work on various Squirrel AI Learning products to solve problems that vary in subjects and difficulty levels.  The dataset resources include user records from learner records store of Squirrel AI Learning system, brainwave data collected by a EEG headset device, and video data captured by web camera while students working on the learning system. The primary aim is to analyze student learning activities, facial expressions, body movements, and brainwave patterns to predict student engagement. This can then be used to improve adaptive learning selection and student learning outcomes. An additional goal is to provide a dataset gathered from truly real-world educational activities versus those from a controlled lab environment. Our dataset can be publicly accessed via the following link. \footnote{\url{https://github.com/RyanH98/SAILData}}

\begin{figure*}[!htb]
\begin{center}
 \includegraphics[scale=0.55]{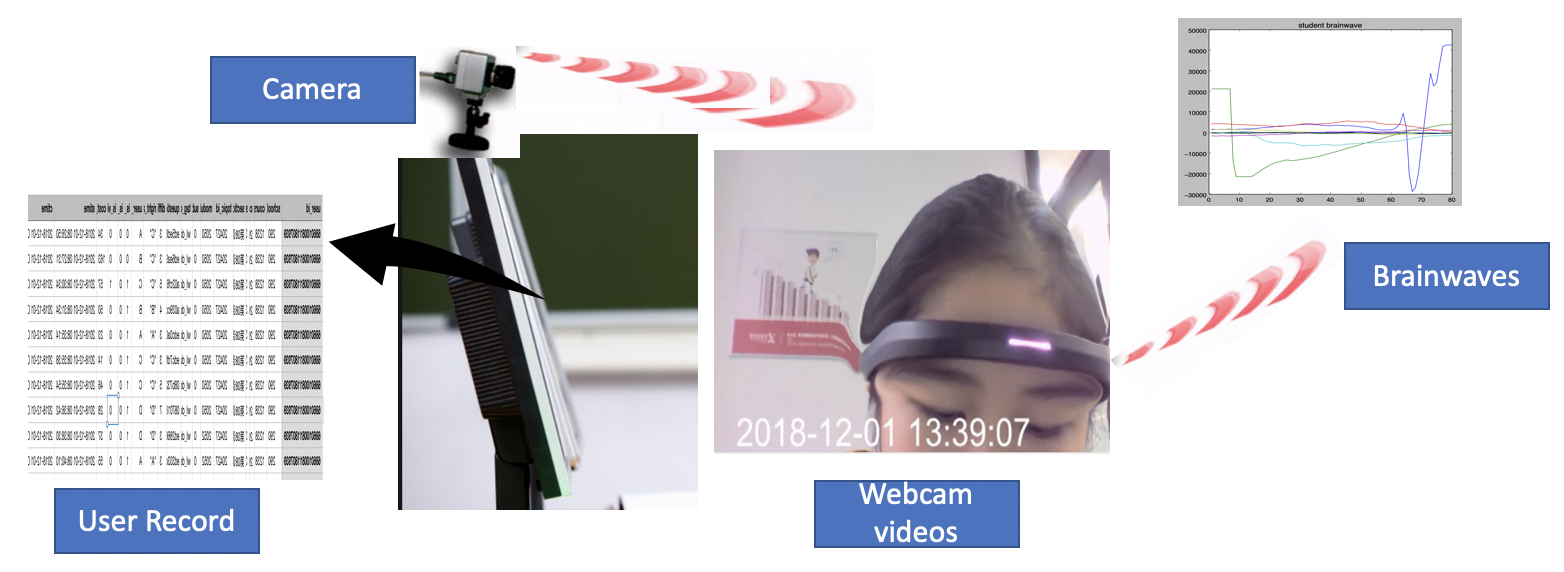}
   \caption{Integration of multimodal data capture using camera and EEG headband. During the question answering process, we collect student's learning record, EEG brainwave signals, and a set of video data.}  
 \end{center}   
\end{figure*}

\section{Method}
\subsection{Participants}

For this dataset, the participants were students from 2 Squirrel AI Learning (SAIL) after-school learning centers in China, with 33 of students coming from school G and 123 students coming from school N. Students worked on 6 Subjects in total. For Chinese and English (Grammar), students came from primary school and middle school. For Math, Physics, Chemistry and English reading, students came from middle school only.

\subsection{Tasks and data collection procedure}

Note that the students in the study go to the SAIL schools after their regular school day. In school G after school tutoring sessions were approximately 1.5 hours long, and in school N they were approximately 2 hours long. In each session, students focus on a particular set of knowledge points, which can be small units of facts, concepts, or skills (called tag\_code in the dataset) for a specific subject. They typically start with a pretest or review then view instructional videos and work on learning and practice questions until the knowledge points have been “mastered”. Pretest questions do not have immediate corrective feedback, but students can view the answer and the explanations of the answer after they complete the pretest. For learning and practice, students can activate varying levels of hints and corrective feedback before submitting their response, and they can view the answer explanation for each question after submitting their response.

After each learning and practice question, SAIL computes and updates each student’s proficiency level on each knowledge point to determine the next question to present. When the proficiency estimates reach a certain threshold, that knowledge point is considered “learned” and will be queued for review at a later time. 

For our study, students wore “brainwave” headsets (manufactured by BrainCo \footnote{\url{ https://www.brainco.tech/}}) while they were studying in the after-school learning session, and they were also asked to have their webcam cameras turned on. Thus, there are three main data sources collected for each student: (i) User records from the SAIL learning record store, (ii) Students brainwave data while learning from the BrainCo headband and stored by FocusEDU platform. (iii) The video data captured by the webcam installed on each computer and stored by Debut video capture software\footnote{\url{ https://www.nchsoftware.com/capture/index.html}}. The whole procedure and data source flows are shown in figure 1. 

\subsubsection{User record}
The user records are collected from Squirrel AI learning's learner record store. They contain all the question level logs of student responses while students are working in the SAIL tutoring sessions for each subject. Each item is a question. Note that students were generally focused on answering questions throughout the session except when they were watching videos to help them learn a particular knowledge component. Students' interactions with the instructional videos were also captured. These include actions such as video  play, fast forward, play back, etc.  

\subsubsection{Brainwaves}
When students work in each learning session, the headsets they are wearing generate three data files: attention, EEG and events. 

The attention data contains a Unix timestamp followed by its attention value. This attention value ranged 0-100 and was generated by the BrainCo devices. Since the data collection occurred in China, this timestamp was then converted to Beijing time so that it can be synchronized to other data sources.

The EEG file contains the raw EEG data. Each row has the timestamp, sequence number, battery level, logging label and EEG array. Each point represents the difference in potential between the EEG reference point and the acquisition point. There are 160 such points in a minute for $uA$.
The vectors in square brackets [ ] are the electrical signal output values of the sensors. These electrical signals can be transformed into frequency domain signals or wave forms (alpha, beta, gamma, etc.) by Fourier transform, and the average energy of each wave band can be calculated.

The event file contains the raw events data. Each data point has a Unix timestamp followed by the device stage. It indicates whether the device is connected or not in the corresponding time. 

\subsubsection{Webcam videos}
Each student’s entire session was also recorded by the webcam on the computer, and this focused on the student’s upper body, including the face. And these video sessions are time stamped so they can be properly synchronized.


\subsection{Multimodal data synchronization and processing}
\subsubsection{User records}
Since we recorded students facial movements by the camera and captured brainwaves by the headset while they working on the after-school session using the SAIL system, we sync these three data sources together by time.
\subsubsection{Brainwave data and synchronization}
There are two steps in brainwave data synchronization. The first step is to match each student with his or her brainwave data, and we used student user id, session time, headset number to do this. The second step is to synchronize each student’s brainwave data with that student’s user record. The algorithm for syncing these two data sources is shown in Algorithm 1. In this algorithm, ctime represents the answer submission time of corresponding question. The main task is to find the right student brainwave triplets while the student working on the question. 
\begin{algorithm}[tb]
\caption{Syncing Brainwave with User Record}
\label{alg:algorithm}
\textbf{Input}: Table of all User record, Table of brainwave data\\
\textbf{Parameter}:user\_id, subject, end\_time\_with\_date, ctime, class\_start\_time, class\_end\_time, brainwave\_file\_path\\
\textbf{Output}: brainwave triplet file paths
\begin{algorithmic}[1] 
\WHILE{user\_id = 52027 and subject = "E" in table Brainwave}
\STATE find the end\_time\_with\_date and class\_start\_time
\WHILE{user\_id = 52027 in English User record}
\STATE find ctime
\IF {ctime is between class\_start\_time and end\_time\_with\_date }
\STATE assign the brainwave\_file\_path to the item which user\_id = 52027 in User record
\ENDIF
\ENDWHILE
\ENDWHILE
\STATE \textbf{return} User record with brainwave file path appended
\end{algorithmic}
\end{algorithm}

\subsubsection{Video processing and synchronization and segmentation}
The webcam video data has the same fields as the brainwave data. We used the same algorithm to sync each student's webcam video data with that student's user record. After the synchronization, each question item in the user record is matched with the original video file path.

After we synchronized the video with the user record, the data were segmented from each session into time phases representing the start and end of each question the student worked on. Each question in the user record has a stime, which means the time the student was given the question, and a ctime, which equals to the submitted time. With this encoding, the total time required by the student to solve each question is simply ctime - stime. And the video piece starting at stime and ending at ctime gives the video data when the student was working on that specific question. 

The extracted question segments have two files associated with each segment: a json file with metadata of the question and an npy (python Numpy) file containing the tracking metadata, and they should be very easy to load into Python. 

1. Question segments filenames name as “school name\_video ids\_segment numbers. 

2. Json file with metadata of the question can be retrieved from user records.

3. File named “npy\_key.md” describing the meanings of the various data point included in the numpy file. 

After the webcam videos were matched with user records, the total video length for different Subjects are computed and are shown in table 1.

After filtering, it contains 2170 segments. Table 2 showed the percentage of valid tracking of the segments for different percentile. We define that fully visible student face as valid tracking. At least 50\% means at least half of the frames have valid tracking information.

\begin{table}
\centering
\begin{tabular}{lll}
\hline
Subject  & total time (seconds)\\
\hline
Chemistry       & 3809     \\
Chinese      & 3835    \\
English   &  25991    \\
Math   &  17012    \\
Physics & 50790\\
\hline 
\end{tabular}
\caption{Total time for different subjects}
\label{tab:plain}
\end{table}

\begin{table}
\centering
\begin{tabular}{lll}
\hline
Percentage of valid segments & Percentage (2170 segments)\\
\hline
At least 50\%   & 61.54\%  \\
At least 70\%   & 49.88\%  \\
100\%  & 18.30\%   \\
\hline 
\end{tabular}
\caption{valid tracking about segments}
\label{tab:plain}
\end{table}


\section{Conclusion and Future Work}
In this paper, we have presented MUTLA, a large-scale multimodal dataset including learning logs, videos, and EEG brainwaves. Because this dataset is collected in a real complex learning environment, we hope that it will greatly facilitate research on measuring and improving student engagement, which can then be used to improve the effectiveness of adaptive learning systems. 

Going forward, we will try to apply machine learning and deep learning techniques ourselves to this multimodal data in the hope better understanding how these complex data sets are related.

\section{Acknowledgement}
Special thanks to the two after-school learning centers staff and students who participated and supported this data collection, and Squirrel AI Learning staff and interns who aided in the data collection, cleaning, and aggregation. The data described and shared here were collected by Squirrel AI Learning, and they also funded this research. SRI International also aided preparation for this dataset.

\bibliographystyle{named}
\bibliography{ijcai19}

\end{document}